\documentclass{emulateapj}

\usepackage{graphicx}
\usepackage{amsmath}
\usepackage{xspace}
\usepackage{hyperref}
\newcommand{\IGRs}{IGR~J17448-3232 }
\newcommand{\IGR}{IGR~J17448-3232}
\newcommand{\inte}{{\em INTEGRAL} }

\def\la{\mathrel{\hbox{\rlap{\hbox{\lower4pt\hbox{$\sim$}}}\hbox{$<$}}} }
\def\ga{\mathrel{\hbox{\rlap{\hbox{\lower4pt\hbox{$\sim$}}}\hbox{$>$}}} }

\def\arcmin{\hbox{$^\prime$} }

\def\flux{ergs~cm$^{-2}$~s$^{-1}$ }
\def\h2{H$_2$}
\def\msol{$M_\odot$}

\def\xmms{{\it XMM-Newton} }
\def\xmm{{\it XMM-Newton}}
\def\chandras{{\it Chandra} }

\def\swifts{{\it Swift} }
\def\swift{{\it Swift}}

\def\integrals{{\it INTEGRAL} }

\begin{document}

\lefthead{Identification of IGR J17448-3232 as a new galaxy cluster}
\righthead{Barri{\`e}re et al.}

\submitted{Accepted by ApJ}

%\title{IGR J17448-3232 field: a blazar and a new nearby galaxy cluster seen through the Galactic center}
\title{Source identification in the IGR J17448-3232 field: discovery of the Scorpius galaxy cluster}

\author{
Nicolas~M. Barri{\`e}re\altaffilmark{1},
John~A. Tomsick\altaffilmark{1},
Daniel~R. Wik\altaffilmark{2},
Sylvain Chaty\altaffilmark{3},
J{\'e}rome Rodriguez\altaffilmark{3}
}

\altaffiltext{1}{Space Sciences Laboratory, University of California, Berkeley, CA 94720, USA.}
\altaffiltext{2}{Astrophysics Science Division, NASA Goddard Space Flight Center, Greenbelt, MD 20771, USA}
\altaffiltext{2}{CEA/DSM-CNRS-Universit{\'e} Paris Diderot, Irfu/Service d'Astrophysique, Centre de Saclay, 91191, Gif-sur-Yvette, France}


\begin{abstract} 
We use a 43-ks \xmms observation to investigate the nature of sources first distinguished by a follow-up \chandras observation of the field surrounding \integrals source \IGR, which includes extended emission and a bright point source previously classified as a blazar. We establish that the extended emission is a heretofore unknown massive galaxy cluster hidden behind the Galactic bulge. The emission-weighted temperature of the cluster within the field of view is 8.8 keV, with parts of the cluster reaching temperatures of up to 12 keV; no cool core is evident.  At a redshift of 0.055, the cluster is somewhat under-luminous relative to the X-ray luminosity-temperature relation, which may be attributable to its dynamical state. We present a preliminary analysis of its properties in this paper. We also confirm that the bright point source is a blazar, and we propose that it is either a flat spectrum radio quasar or a low-frequency peaked BL Lac object. We find four other fainter sources in the field, which we study and tentatively identify. Only one, which we propose is a foreground Galactic X-ray binary, is hard enough to contribute to \IGR, but it is too faint to be significant. We thus determine that \IGRs is in fact the galaxy cluster up to $\approx45$ keV and the blazar beyond.
\end{abstract}





\section{Introduction}

%short intro on IGR sources
Since its launch in 2002, the INTErnational Gamma Ray Astrophysical Laboratory \citep[INTEGRAL;][]{winkler.2003ly} has been key to discovering non-thermal hard X-ray sources (``IGR'' sources), thanks in particular to the large field of view of IBIS \citep[$29^{\circ} \times 29^{\circ}$ partially coded;][]{ubertini.2003bf}. Determining the nature of these new sources often requires the identification of optical and infrared (IR) counterparts. However the position resolution of IBIS ($1'$ -- $5'$ for sources detected in the 20--40 keV range) is generally too coarse to match with a unique counterpart, especially in the Galactic center.

%Intro on this particular IGR source
\IGRs was  reported in the third and fourth IBIS/ISGRI soft gamma-ray survey catalogues \citep{bird.2007ys, bird.2010fk}. This field was observed with {\it Chandra} for 5~ks in 2008 November to refine the location and identify the IGR source \citep{tomsick.2009uq}. Two possible counterparts were uncovered: an extended source, CXOU~J174453.4-323254, and a point source with a hard spectrum, CXOU~J174437.3-323222. Given the proximity with the Galactic center (l=356.8$^{\circ}$, b=-1.8$^{\circ}$), it was proposed that the sources are a supernova remnant with its pulsar moving away. Using new and archival observations, \citet{curran.2011fk} were able to build a spectral energy distribution (SED) for the point source and determined that it corresponds to a blazar viewed through the Milky Way. 

%This paper + outlines
The \xmms observatory \citep{jansen.2001vn} has three co-aligned X-ray telescopes, with the three EPIC spectro-imagers at their focus: two use MOS CCDs \citep{turner.2001kx} and one uses a PN CCD \citep{struder.2001ys}. We present here our \xmms observation of the \IGRs field that allows us to identify the extended emission as a previously unknown massive galaxy cluster in the Scorpius constellation. We also provide a measurement of the blazar candidate with better statistical quality than were previously available. Section \ref{sec:data} presents the observation and the tools used to reduce the data, and our new image of the field is shown in section \ref{sec:map}. We then focus on the galaxy cluster in section \ref{sec:GC} and on the blazar in section \ref{sec:blazar}. Finally we investigate the other sources in the field in section \ref{sec:othersrc}, before concluding. We assume a flat cosmology with $\Omega_M = 0.23$ and $H_0 = 70$~km~s$^{-1}$~Mpc$^{-1}$. Unless otherwise stated, all uncertainties are given at the 90\% confidence level.

\begin{figure}[t]
\begin{center}
\includegraphics[width=0.45\textwidth]{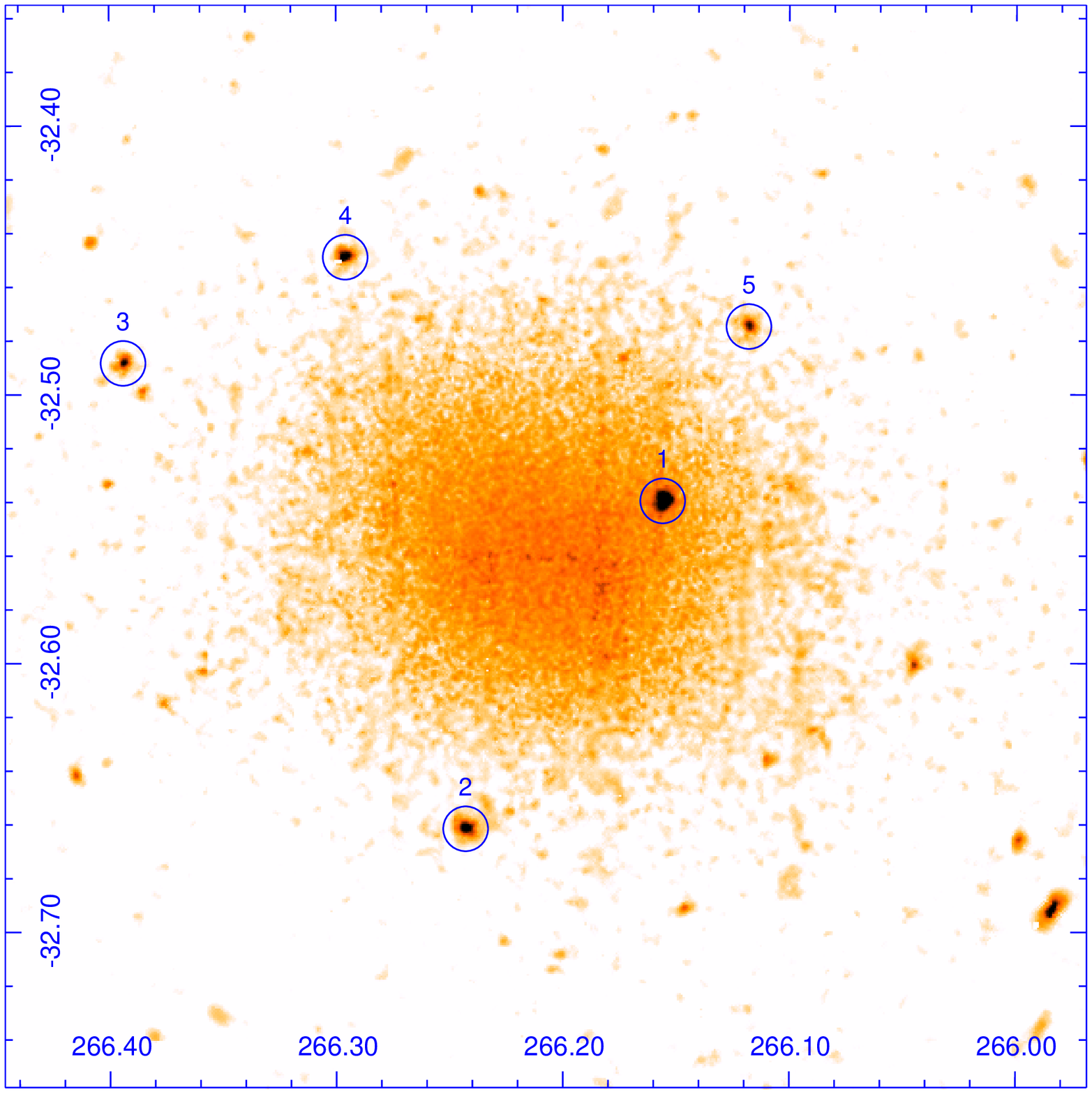}
\caption{Combined (MOS1, MOS2, pn) background-subtracted and exposure-corrected image of the \IGRs field, in the 0.4--7.2~keV energy range. The image spans $24' \times 24'$ and is oriented with equatorial coordinates, the North being to the top, and the East to the left, with logarithmic color scale. The galaxy cluster lies in the middle, and the 5 blue circles ($30''$ radius) show the sources listed in Table \ref{tab:pointsrc} with their corresponding ID number.
\label{fig:image}}
\end{center}
\end{figure}

%%%%%%%%%%%%%%%%%%%%%%%%%%%%%%%%%%%%%%%%%
\section{Dataset and analysis tools}
\label{sec:data}
We use a 43.9-ks \xmms observation that started on 2012 February 26, 03:59:17 UT (ID 0672260101). The data analysis is split between the point sources and the extended emission. The data for the point sources was reduced using XMM SAS v13.5.0, while the extended emission data was reduced using the Extended Source Analysis Software (XMM-ESAS\footnote{http://heasarc.gsfc.nasa.gov/docs/xmm/xmmhp\_xmmesas.html}) package from SAS version 13.0.0. \citet{snowden.2008fk} introduced the analysis of galaxy clusters with ESAS, and it has then been expanded to include the EPIC pn data \citep[e.g.,][]{bulbul.2012fk}.

The spectral analysis is done with XSPEC \citep{arnaud.1996uq}, using \citet{verner.1996kx} photoelectric cross-sections and \citet{wilms.2000vn} abundances, and $\chi^2$ statistics.  %Errors are quoted at the 90\% confidence limit, unless otherwise noted.

%\cite{snowden.2008fk}: XMM ESAS package and catalogue of galaxy cluster observed by XMM.
%and a 4.7-ks {\it Chandra} observation that started on UT 2008 November 2 at 02:08:53 (ID 9059) using the Advanced CCD Imaging Spectrometer \cite[ACIS][]{garmire.2003uq}

%%%%%%%%%%%%%%%%%%%%%%%%%%%%%%%%%%%%%%%%%
\section{Overview of the field: source detection}
\label{sec:map}

Using the XMM-ESAS package, we first filtered out the flaring background events, and created a model of the quiescent particle background (QPB) for each camera over the 0.4--7.2~keV range. The three images were then background-subtracted and exposure-corrected before being combined. The resulting image was adaptively smoothed to reach 30 counts per circle (Figure \ref{fig:image}). The extended source CXOU~J174453.4-323254 is clearly visible at the center of the image, as well as a handful of point sources. We ran the task {\it cheese} that invokes {\it emldetect} to detect and localize the sources present in the field. Using a maximum likelihood threshold of $10^3$, we find five point sources, as listed in Table \ref{tab:pointsrc}, the brightest one being CXOU~J174437.3-323222 (\#1). Several other point sources are detected but fall below our threshold for further study. The source at the bottom right corner of the image is outside the field of view of MOS1 and MOS2, and is $15.8'$ off-axis in pn, which explains why it appears elongated. We will not consider this source further.

As the \IGRs field was observed in 2008 by \chandras (obsID 9059), we found good match between the sources circled in Figure \ref{fig:image} and the \chandras ones reported in \citet{tomsick.2009uq}. We used the \chandras positions to look for counterparts for these sources in section \ref{sec:othersrc}.

\begin{deluxetable*}{cccccc}
\tabletypesize{\small}
\tablecaption{ID, name and position of the sources detected in the {\it XMM-Newton} observation (Equatorial coordinates, J2000). Source \#0 is the extended emission. \label{tab:pointsrc}}
\tablecolumns{6}
%\tablewidth{0.85\textwidth}
\tablehead{\colhead{ID} & \colhead{Name} & \colhead{R.A.}   &  \colhead{Decl.}  & \colhead{Err ($''$)} & \colhead{ $\Delta_{\rm pos}$ ($''$)} }
\startdata
0 & CXOU J174453.4-323254  &  $17^h44^m49^s.73$  & $-32^{\circ}33' 25''.9$ 	&  	\nodata	& 	\nodata   \\
1 & CXOU J174437.3-323222  &  $17^h44^m37^s.44$  & $-32^{\circ}32' 22''.6$ 	&  0.37		& 1.33 \\  % Chandra: 17 44 37.34	-32 32 23.0	
2 & CXOU J174458.2-323940 & $17^h44^m58^s.32$  & $-32^{\circ}39' 41''.8$   	& 1.25 	& 1.25 \\ %Chandra: 17 44 58.29	-32 39 40.4  
3 & CXOU J174534.6-322917 & $17^h45^m34^s.56$  & $-32^{\circ}29' 17''.9$		& 4.15 	& 1.30 \\ %Chandra:17 45 34.66	-32 29 17.6 
4 & CXOU J174510.9-322655 & $17^h45^m11^s.04$  & $-32^{\circ}26' 56''.0$  	& 0.75 	& 1.06 \\ %Chandra: 17 45 10.96	-32 26 55.7
5 & CXOU J174428.4-322828 & $17^h44^m28^s.32$  &  $-32^{\circ}28' 28''.9$ 	& 2.04 	& 1.28 \\ % Chandra: 17 44 28.42	-32 28 28.7 
\enddata
%\vspace{-0.8cm}
\tablecomments{The Err column gives the statistical error at the 95\% confidence level of the \chandras position, calculated using Eq. 5 of \citet{hong.2005fk}. It does not account for the 0.64$''$ (90\% confidence level) systematic uncertainty on the
\chandras pointing. The $\Delta_{\rm pos}$ column gives the angular distance between the \xmms position derived form this observation and the \chandras position (quoted in this table). }
\end{deluxetable*}

%%%%%%%%%%%%%%%%%%%%%%%%%%%%%%%%%%%%%%%%%
\section{CXOU J174453.4-323254: a new galaxy cluster}
\label{sec:GC}
 
\subsection{Data filtering and Background Modeling}

Data are first cleaned using standard SAS routines, reducing the event list to 43~ks of on-target exposure. Time intervals affected by proton flaring events are then filtered out, leaving 25.4~ks, 26.9~ks, and 16.8~ks of exposure in MOS1, MOS2, and pn, respectively. 
We also exclude CCD 6 of MOS1 and CCD 5 of MOS2 from our analysis since
those CCDs were flagged as ``anomalous,'' which indicates heightened and non-standard
particle background. Point sources with fluxes larger than $5 \times 10^{-15}$ \flux\ are
identified and excluded out to a radius where the surface brightness of the PSF
is 50\% of the local background surface brightness. Images and spectra are then extracted. The ESAS allows estimating the quiescent particle
background using a database of observations done with closed filter wheel that are matched to data from the unexposed corners of the cameras. The remaining background components (cosmic fore- and backgrounds, solar wind charge exchange emission, residual soft proton contamination, and instrumental lines) cannot be subtracted so they are modeled and fit to the data. No charge exchange line emission is apparent in these data.

The cosmic (extragalactic) background (CXB) is modeled as a simple power law with
photon index 1.46 and assumed total flux equivalent to that found by \citet{lumb.2002uq}.
Because bright sources have been removed, the residual CXB in our spectra is only a fraction of the total flux. The measured 0.5--8~keV band $\log N - \log S$ of \citet{kim.2007kx} is used to calculate the total flux of sources above our threshold, which amounts
to 75\% of the total CXB flux. We assume the remaining CXB sources are spread uniformly within our regions and fix their contribution at 25\% of the total CXB flux. 
% is absorbed by the same $N_{\rm H}$ column as the cluster, a higher

The cosmic (Galactic) foreground consists of unabsorbed emission from the local hot bubble ($kT \sim 0.1$~keV) and absorbed emission from the Galactic halo (modeled as a two-phase plasma with $kT_1 \sim 0.1$~keV and $kT_2 \sim 0.6$~keV) and the Galactic plane (called the Galactic Ridge X-ray Emission, or GRXE). Although the GRXE may consist largely of point sources \citep[e.g.,][]{revnivtsev.2006vn},
we assume negligible GRXE flux has been eliminated during point source removal.
We model its spectrum following \citet{valinia.1998zr} for their Region 3, which corresponds to
the southern Galactic longitudes where our observation lies.
Given the spatial variation of the GRXE, its actual flux may be higher or lower,
but we simply fix it during fits. It accounts for roughly 20\% of the total observed 0.5--2~keV flux in our observation, but any fluctuations from around this value can easily be accommodated by the
thermal Galactic components, which have free normalizations and temperatures.
We fix the value of $N_{\rm H}$ to the LAB column density of $7 \times 10^{21}$~cm$^{-2}$
\citep{kalberla.2005uq}.

\subsection{Spectral modeling}

For the spectral modeling, we follow the method established by \citet{snowden.2008fk}. The model $M$ has two main components: $M_{\rm resp}$, which is folded through the nominal RMFs and ARFs, and $M_{\rm diag}$, which is only folded through a diagonal response. $M_{\rm diag}$ represents the residual soft proton contamination, and $M_{\rm resp}$
everything else: the instrumental lines ($I_{\rm lines}$), the soft emission from the
local hot bubble ($G_{\rm LHB}$), the thermal Galactic disk emission and the GXRE ($G_{\rm halo}$), the extragalactic cosmic X-ray background ($B_{\rm CXB}$), and finally the intracluster medium emission ($S_{\rm cluster}$).

In XSPEC, these components are made of the following models:
\begin{align*}
&I_{\rm lines} = \sum {\tt gauss} \\
&G_{\rm LHB} = {\tt apec} \\
&G_{\rm halo} = {\tt apec} + {\tt apec} + ({\tt apec} + {\tt powerlaw})_{\rm GRXE} \\
&B_{\rm CXB} = {\tt powerlaw} \\
&S_{\rm cluster} = {\tt apec}
\end{align*}
These components are modulated by total Galactic absorption $N_{\rm H, 2}$,
foreground Galactic absorption $N_{\rm H, 1}$ (likely between us and the 3~kpc
molecular ring), and normalization constants that account for imperfect
cross-calibration between the three detectors ($C_{\rm det}$) and the differences
in solid angle between each annular region ($C_{\rm area}$). Thus, $M_{\rm resp}$ can be developed as
\begin{multline*}
M_{\rm resp} = I_{\rm lines} + C_{\rm det} C_{\rm area} [G_{\rm LHB} +  N_{\rm H, 1} G_{\rm halo} + \\
N_{\rm H, 2} (B_{\rm CXB} + S_{\rm cluster})],
\end{multline*}
which, gives in pseudo-XSPEC format:
\begin{multline*}
M_{\rm resp} = {\tt gauss+gauss+gauss+gauss+gauss+} \\ 
{\tt gauss + gauss + const \times const \times}  \\ 
{\tt (apec+wabs \times (apec+apec+apec+po)} \\
+ {\tt wabs \times (po+apec) ).}
\end{multline*}

The normalization of the soft proton component is determined for each detector separately, tied together between the annuli. The normalization factor is composed of a constant scaling with the annuli areas multiplied by an additional constant to account for the non-uniform distribution across the detector plane. $M_{\rm diag}$ is expressed as
\begin{equation*}
M_{\rm diag} =  C^\prime_{\rm area} I_{\rm SP},
\end{equation*}
which translates in pseudo-XSPEC format as
\begin{equation*}
M_{\rm diag} =  {\tt const} \times {\tt const} \times {\tt powerlaw}.
\end{equation*}

Because this observation is so near the Galactic plane, molecular hydrogen \h2 is
likely to contribute significantly to the total absorption of the extragalactic CXB and
cluster emission. The \h2 column density may also absorb some of the Galactic emission (GRXE and Galactic halo emission), but for
simplicity we assume it lies in front of any \h2. This assumption might slightly bias our characterization of these components, but as
long as the emission is accounted for the values of the model being used are unimportant
since we are not studying diffuse Galactic X-ray emission and are not interested in modeling each 
component perfectly as long as the total flux is accurately accounted for.
To estimate the \h2 column along our line of sight, we use the velocity-integrated CO 
brightness temperature $W_{\rm CO}$ and its conversion to \h2 column density
$N_{H_2}$ at this location in the Galaxy from \citet{dame.2001ve}.
The total Hydrogen column is then
\begin{equation}
N_{\rm H} = N_{HI} + 2N_{H_2} = N_{HI} + 2(N_{H_2}/W_{\rm CO})W_{\rm CO} \,
\end{equation}
where $N_{H_2}/W_{\rm CO} = 3 \times 10^{20}$~cm$^{-2}$~K$^{-1}$~(km/s)$^{-1}$
and $W_{\rm CO} \sim 20$~K~km/s, or $N_{\rm H} \sim 1.9 \times 10^{22}$~cm$^{-2}$.
This column is nearly three times what would be expected from atomic Hydrogen alone
but is in far better agreement with the data.
In practice, we allow the $N_{\rm H}$ parameter operating on the cluster emission to be
free since it is very well constrained by the data and find it to be only $\approx 14$\% higher than
the estimated value from atomic and molecular Hydrogen.

\begin{figure}[t]
\noindent\null\includegraphics[width=0.45\textwidth]{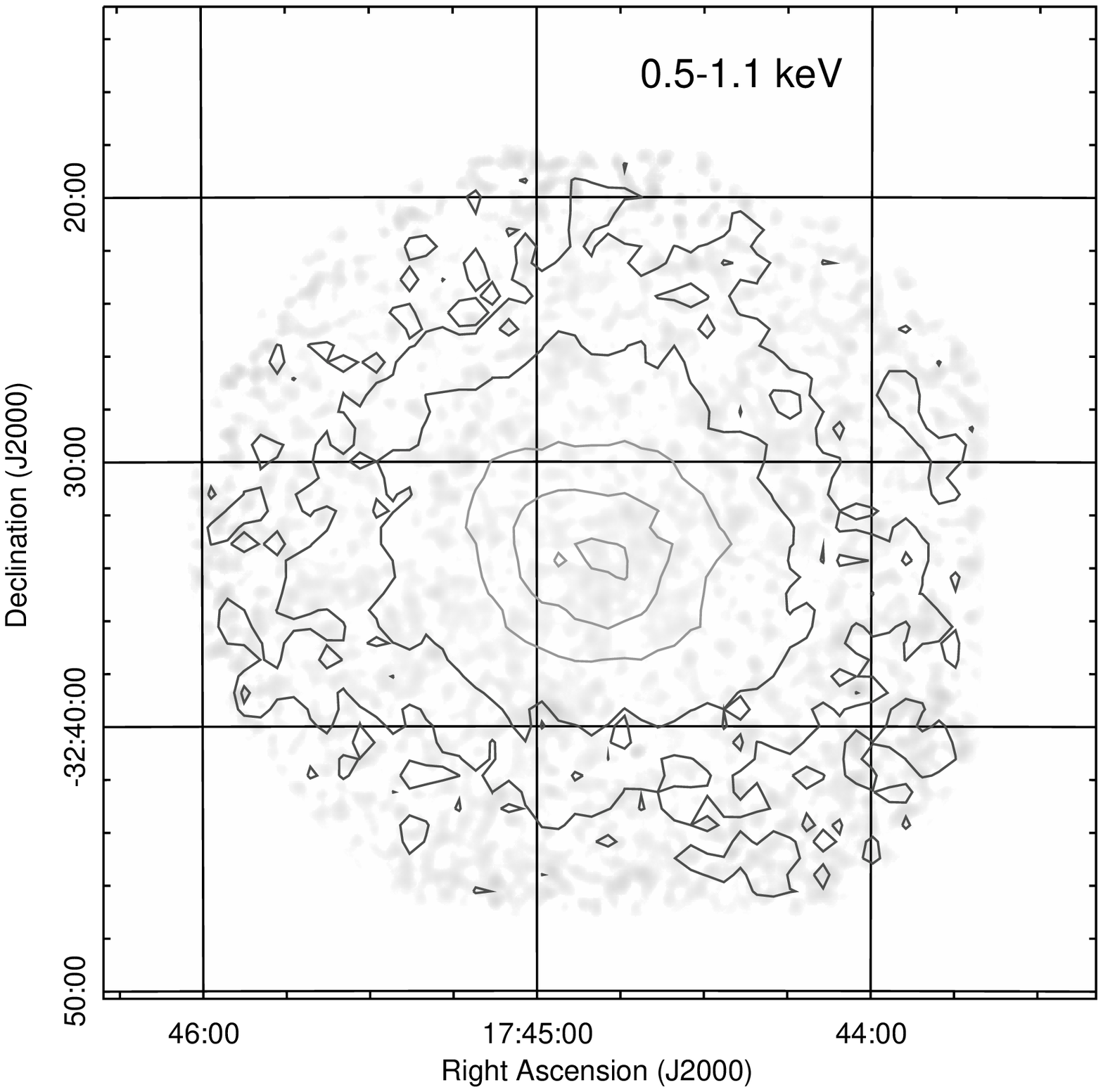} \hfill
\includegraphics[width=0.45\textwidth]{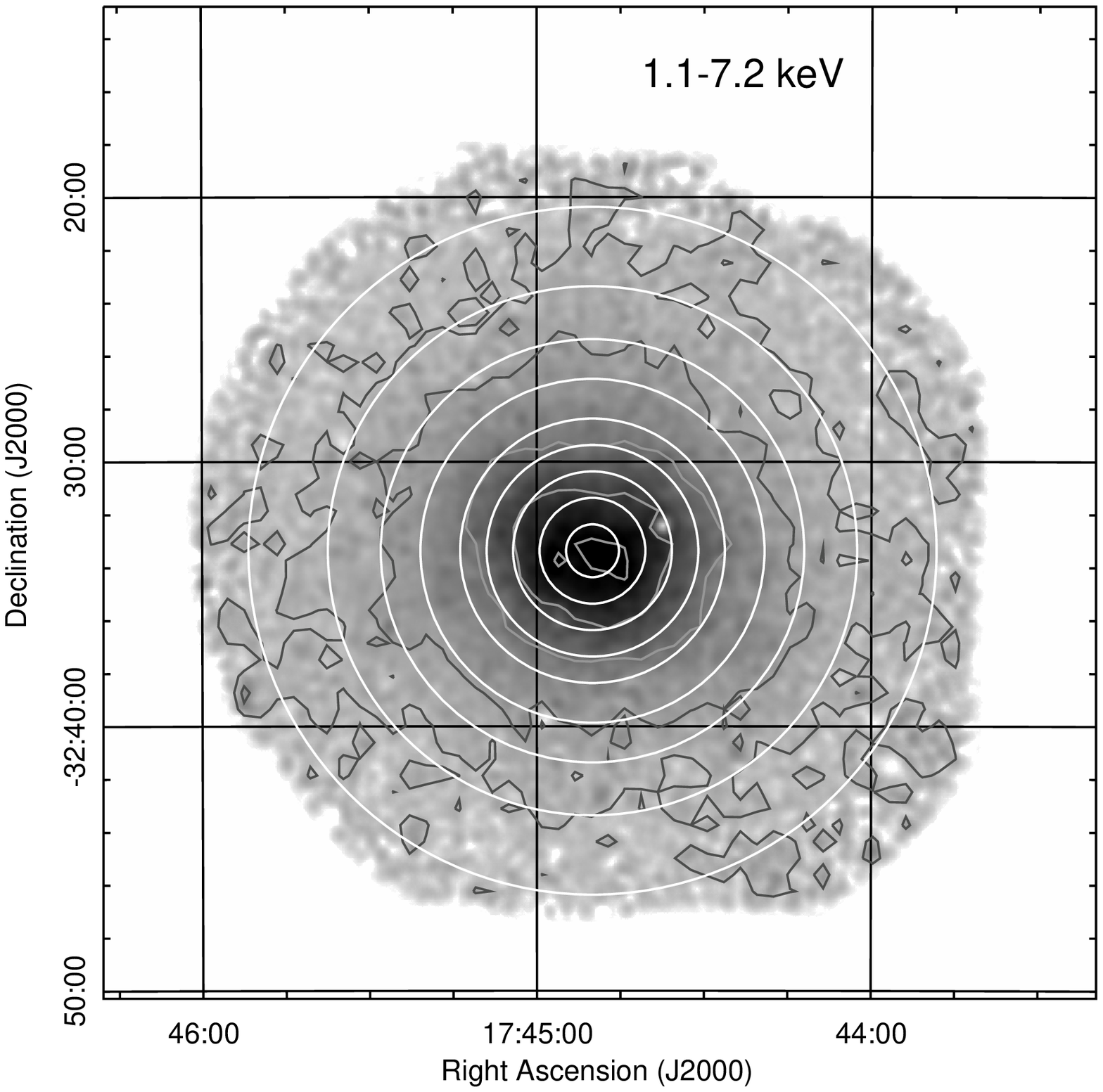}
\caption{Combined, foreground and background-subtracted, and exposure-corrected
EPIC MOS and pn images in the 0.5--1.1~keV (left panel) and 1.1--7.2~keV (right panel)
energy bands.
Each image is scaled from 0 (white) to 120 (black) cts~s$^{-1}$~deg$^{-2}$ and
smoothed with a Gaussian kernel of 22.5\arcmin.
The gray contours in both panels follow the surface brightness of the 1.1--7.2~keV image
with a square-root scaling between 10 and 120 cts~s$^{-1}$~deg$^{-2}$.
In the right panel, circular annuli indicate the 9 regions from which spectra are extracted.
While essentially all of the cluster emission has been obscured by our Galaxy in the soft
band image, at harder energies the cluster is detected out to the edge of the FOV.
\label{fig:clusterimages}}
\end{figure}

\begin{figure}[t]
\includegraphics[width=0.45\textwidth]{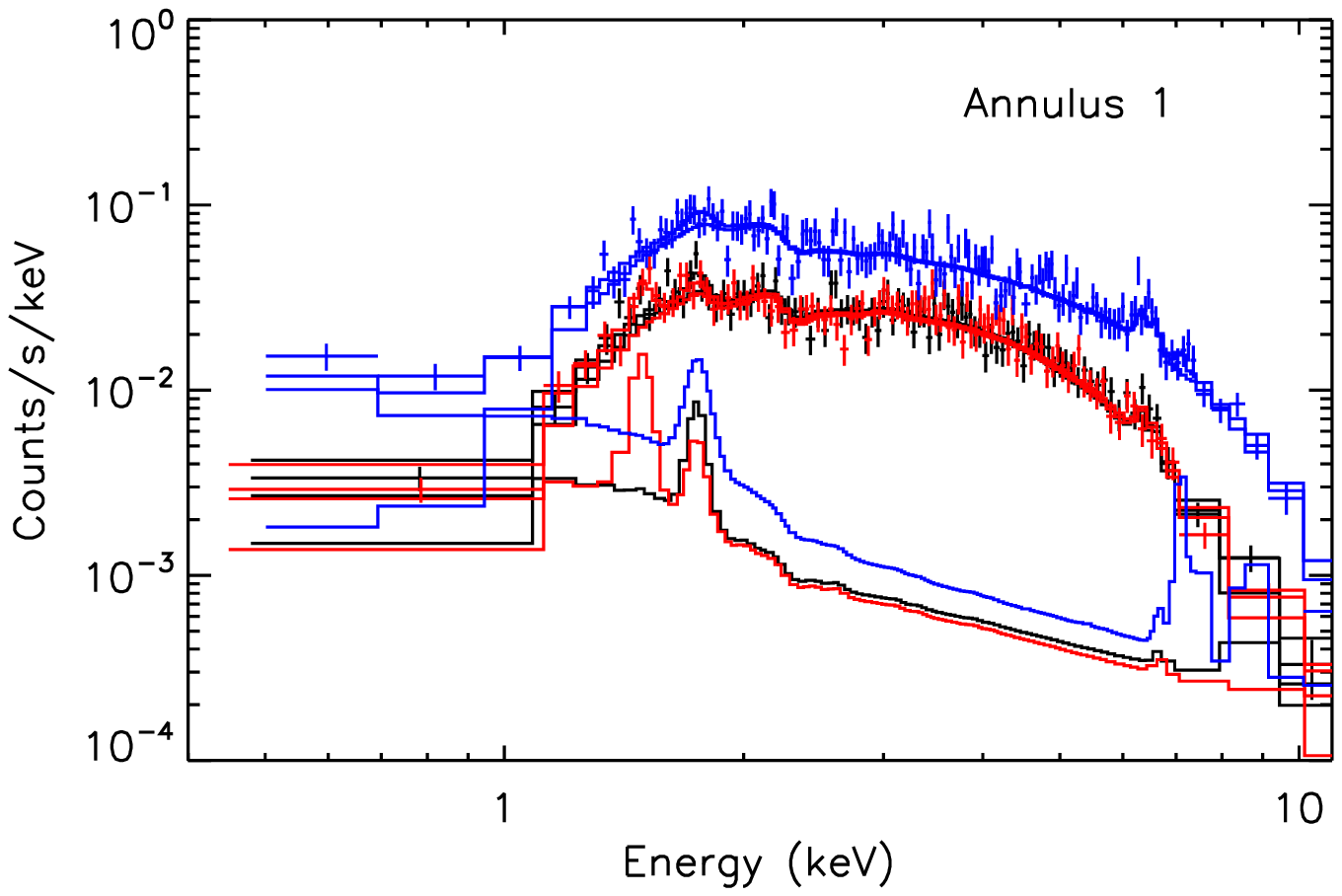} \hfill\null
\noindent\null\hfill\includegraphics[width=0.45\textwidth]{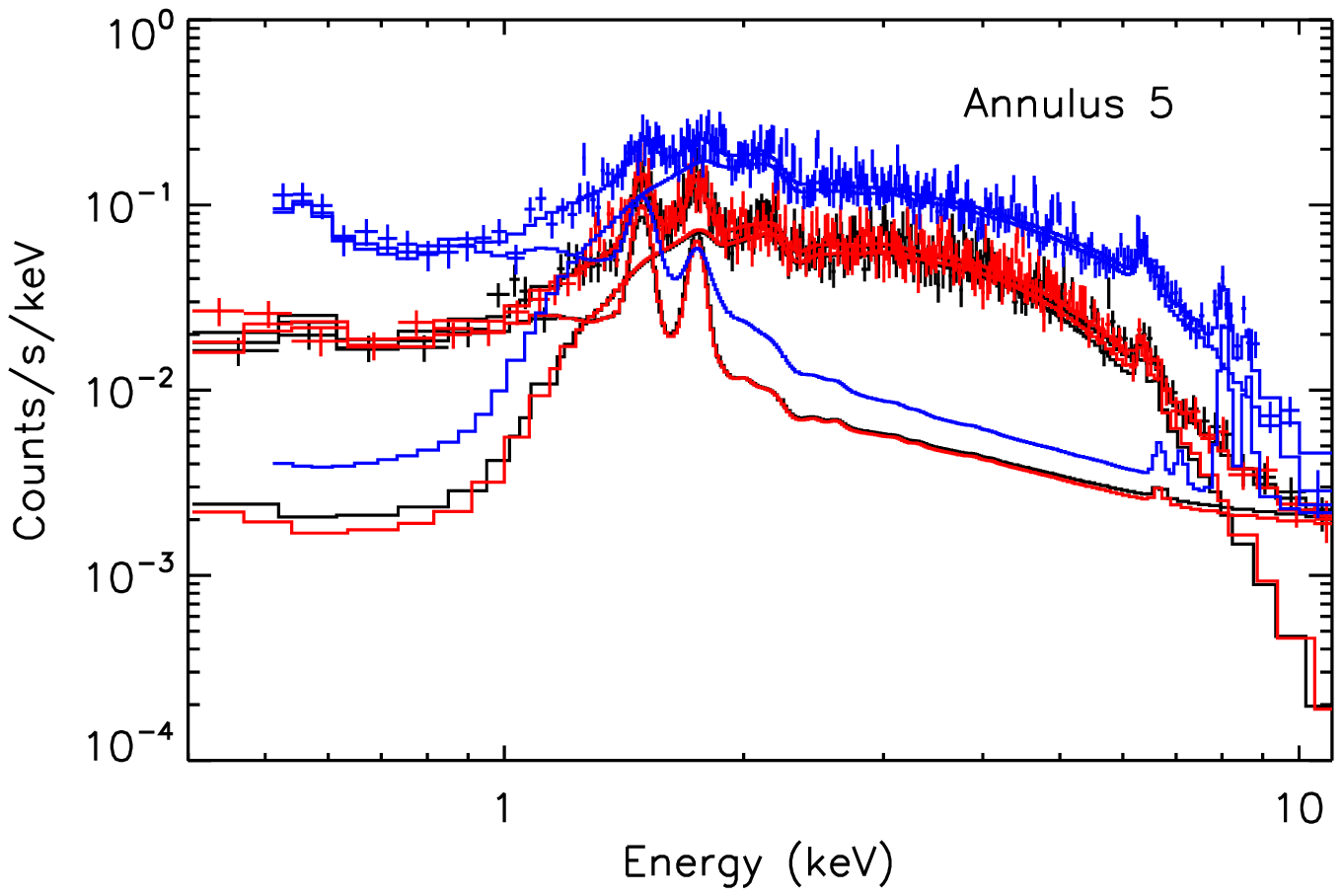} \hfill\null
\noindent\null\hfill\includegraphics[width=0.45\textwidth]{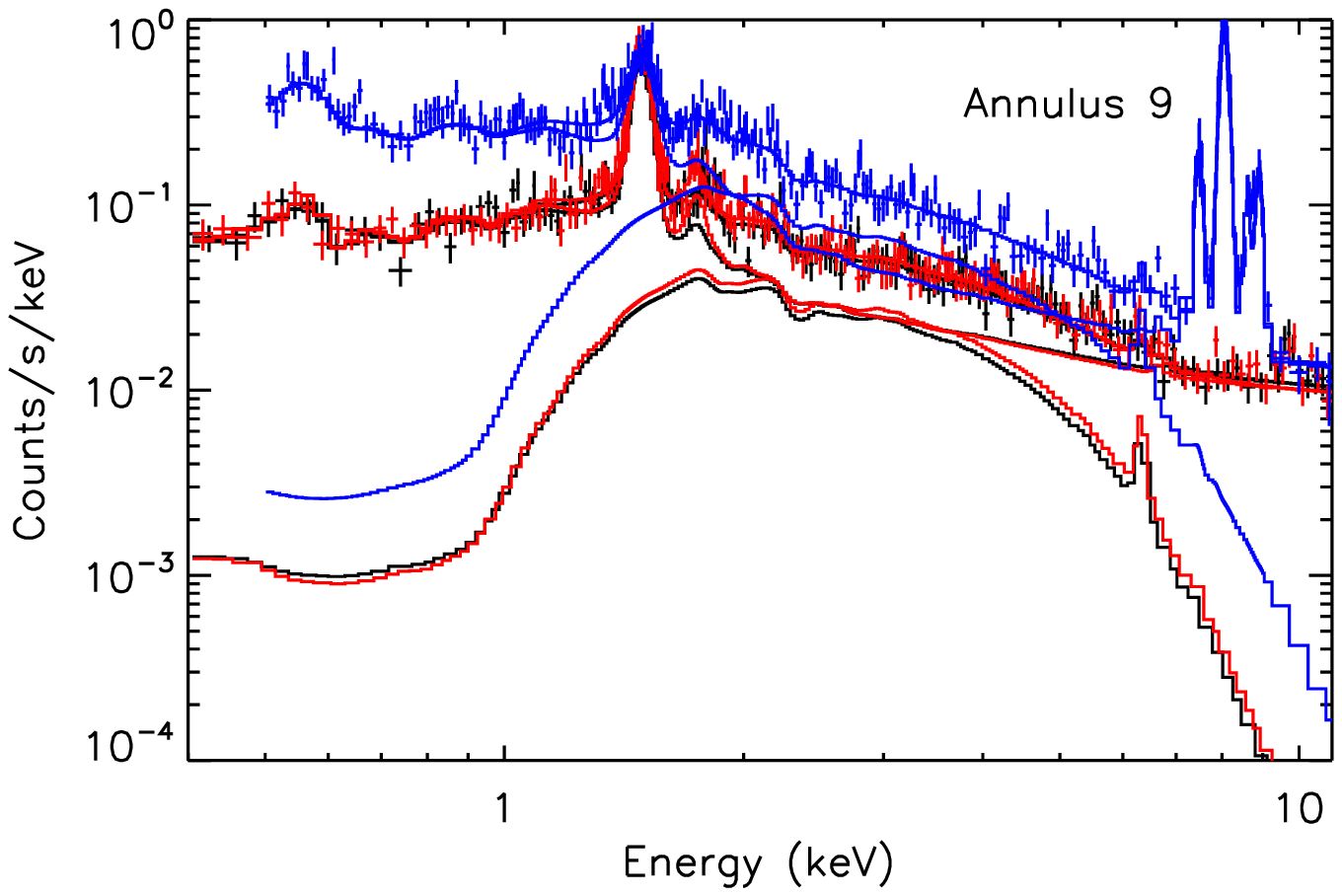} \hfill\null
\caption{MOS 1 (black), MOS 2 (red), and pn (blue) spectra from the innermost (top), middle (middle), and
outermost (bottom) annuli. The best fit models are decomposed into two components: (1) the cluster emission and (2) the sum of Galactic emission, soft protons, and instrumental line background (which are considered as background). The cluster component dominates the 2--6 keV range in all but annulus 9, and single temperature fits are largely adequate to describe this emission,
although a small spread in temperature components distributed throughout an annulus
would still allow a good fit with a single temperature model given the quality of these data.
\label{fig:clusterspec}}
\end{figure}

\begin{figure}[t]
\begin{center}
\includegraphics[width=0.45\textwidth]{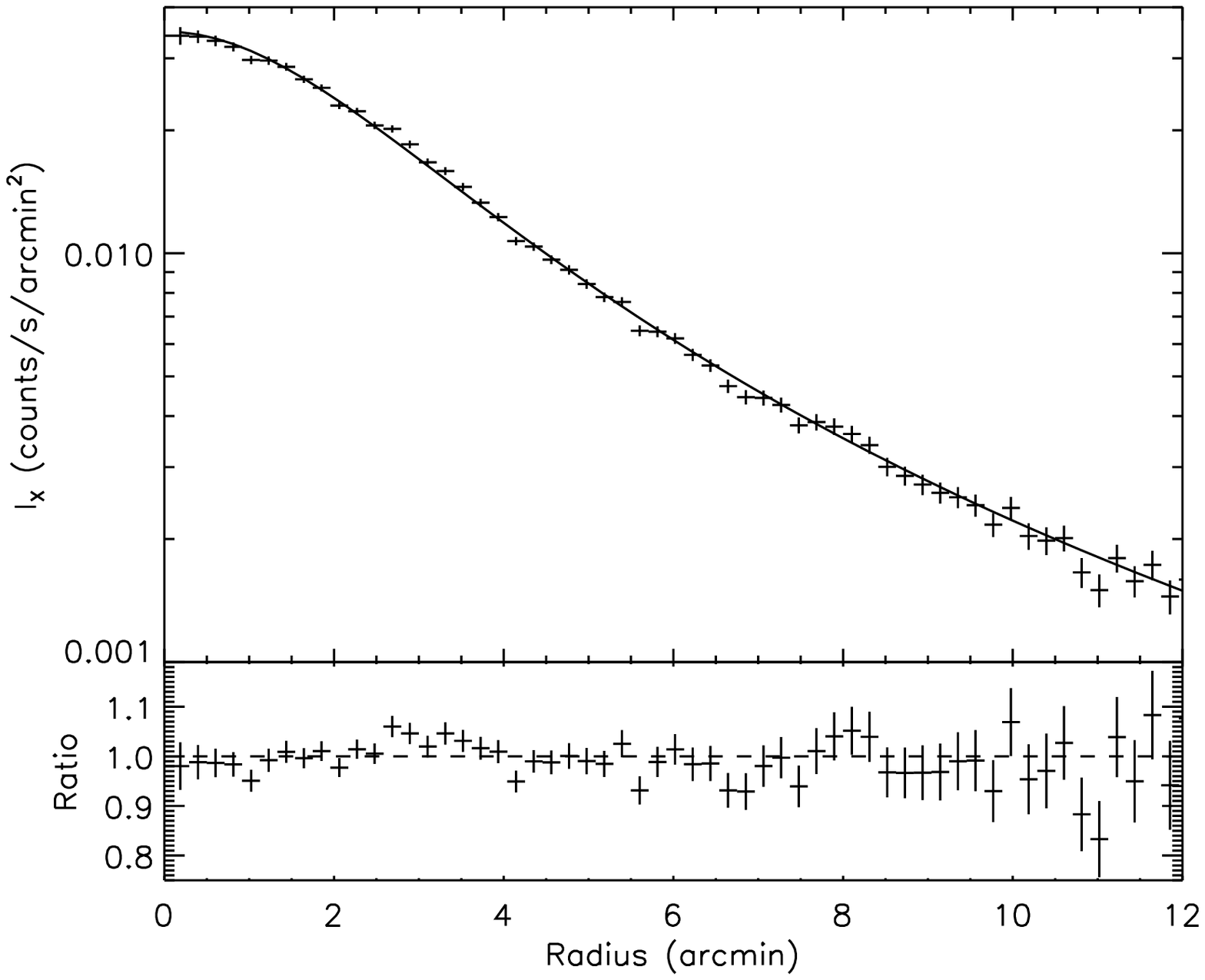}
\caption{Surface brightness profile of the cluster fit with an isothermal $\beta$-model profile. See text for details.
\label{fig:sbprof}}
\end{center}
\end{figure}

%%%%%%%%%%%%%%%%%%%
\subsection{Cluster Properties}

To illustrate the effect of the high absorbing column, combined EPIC images in
the 0.5--1.1~keV and 1.1--7.2~keV bands are shown in Figure~\ref{fig:clusterimages}.
Essentially no cluster emission is visible in the soft band image (left panel)
corresponding to where it is brightest in the hard band image, indicated by the
gray contours in both panels that follow the hard band surface brightness.
We extract spectra from the 9 concentric annuli shown in the right panel of the figure
and simultaneously fit them (along with a spectrum from the ROSAT all sky survey to further constrain the local hot bubble and Galactic foreground contributions) with the model defined above.

Grouping the spectra by at least 30 counts/bin, we obtain an excellent fit with $\chi^2_\nu$ = 1.00 (9576 degrees of freedom, dof).
The absorption ($N_{\rm H} = (2.24 \pm 0.03) \times 10^{22}$~cm$^{-2}$) and Galactic emission are assumed to be uniform across the FOV. Based on constraints provided largely by the Fe-K line complex\footnote{The {\tt apec} model produces an emission spectrum from collisionally-ionized diffuse gas calculated using the ATOMDB code v2.0.2 \url{http://atomdb.org/}. In this model, the redshift is a free parameter. The Fe $K\alpha$ lines are the brightest lines in this model and provide the main constraint to determine the redshift of the emitting plasma}, we measure a redshift $z = 0.055 \pm 0.001$. 
The cluster properties within each annulus are given in Table~\ref{tab:cluster}.
%(In lieu of a table one could plot $kT$ and abundance, as in Figure~\ref{fig:annprof}, but I prefer the table over the figure.)
Although non-unitary cross-calibration constants between the three detector planes
are not necessary to obtain a good fit, a slightly better fit results from allowing them
to be free parameters; the MOS2 and pn data are found to be 
1.4\% brighter and 5.2\% fainter than the MOS1 data, respectively.
The spectral parameters evolve smoothly from the inner annulus to the outer one, so we only show the spectral fit  for the inner, middle and outer ones in Figure~\ref{fig:clusterspec}.

Strikingly, the central region is very hot ($kT \gtrsim 10$~keV), implying
a very massive ($M_{\rm vir} > 10^{15}$~\msol) galaxy cluster heretofore hidden by
our Galaxy. The emission-weighted (2--10~keV) temperature, which we take as a proxy for the
virial temperature, is $kT_{\rm vir} = 8.8$~keV.
However, the bolometric X-ray luminosity within the FOV is 
$L_{X,bol} = 9.6 \times 10^{44}$~erg~s$^{-1}$, or about three times less than
the expected luminosity for a low redshift cluster with this temperature
\citep[e.g.,][]{maughan.2012qf}, although still within the scatter of the relation
for non-cool core clusters.

We estimate that 10--20\% of the total luminosity is lost beyond the \xmms FOV (covering roughly out to $R_{2500}$\footnote{$R_{2500}$ is defined as the radius enclosing mean overdensity of $2500 \rho_{\rm cr}$, where $\rho_{\rm cr}$ is the critical mean density of the Universe, which can be defined in terms of the Hubble function $H(z)$: $\rho_{\rm cr}(z) = 3H(z)^2 / (8\pi\, G)$.}) given the inferred mass and measured redshift. This estimate for the lost luminosity is consistent with the extrapolation of a $\beta$-model fit to the surface brightness profile, given in Figure~\ref{fig:sbprof}. Although the reduced $\chi^2$ is $ \approx 1.5$, the quality of the fit should be considered adequate given the lack of circular symmetry apparent in Figure~\ref{fig:clusterimages}.
For the isothermal $\beta$-model profile \citep{cavaliere.1976dq},
\begin{equation}
I_X = I_0 [ 1 + (R/R_c)^2 ]^{-3\beta+0.5}, \,
\end{equation}
we find a scale radius $R_c = 3.27 \pm 0.06$~arcmin, $\beta = 0.56 \pm 0.01$,
and central surface brightness $I_0 = 0.0349 \pm 0.0004$~cts~s$^{-1}$~arcmin$^{-2}$.
This value of $\beta$ and scale radius (210~$h_{70}^{-1}$~kpc) 
is fairly typical of nearby, hot galaxy clusters without cool cores \citep[e.g.,][]{reiprich.2002cr}.

\begin{deluxetable*}{cccccc}
\tablecaption{Cluster best-fit values for the 9 annuli
\label{tab:cluster}}
\tablehead{
 & $R_{in}$ & $R_{out}$ & $kT$ & abund.\ & Norm.\tablenotemark{a} \\
Annulus & (arcmin) & (arcmin) & (keV) & (rel.\ to solar) & ($10^{-4}$ cm$^{-5}$ arcmin$^{-2}$)}
\startdata
1 & \phn0.0 & \phn1.0 & $11.0^{+1.5}_{-1.0}$ & $0.26^{+0.10}_{-0.09}$ & $7.96^{+0.24}_{-0.17}$ \\
2 & \phn1.0 & \phn2.0 & $11.9^{+1.0}_{-0.9}$ & $0.34^{+0.07}_{-0.06}$ & $6.85 \pm 0.13$ \\
3 & \phn2.0 & \phn3.0 & $11.6^{+1.0}_{-0.8}$ & $0.31^{+0.06}_{-0.04}$ & $5.14^{+0.09}_{-0.10}$ \\
4 & \phn3.0 & \phn4.0 & $10.1 \pm 0.7$ & $0.27 \pm 0.05$ & $3.41^{+0.08}_{-0.04}$ \\
5 & \phn4.0 & \phn5.0 & $\phn9.6^{+0.9}_{-0.8}$ & $0.24^{+0.06}_{-0.05}$ & $2.26^{+0.06}_{-0.05}$ \\
6 & \phn5.0 & \phn6.5 & $\phn7.8 \pm 0.5$ & $0.23^{+0.05}_{-0.04}$ & $1.55^{+0.05}_{-0.04}$ \\
7 & \phn6.5 & \phn8.0 & $\phn6.7^{+0.9}_{-0.5}$ & $0.26 \pm 0.06$ & $1.02 \pm 0.04$ \\
8 & \phn8.0 & 10.0 & $\phn5.7 \pm 0.6$ & $0.14^{+0.07}_{-0.06}$ & $0.69^{+0.04}_{-0.03}$ \\
9 & 10.0 & 13.0 & $\phn3.9^{+0.5}_{-0.4}$ & $0.20^{+0.12}_{-0.09}$ & $0.41 \pm 0.03$
\enddata
\tablenotetext{a}{Normalization of the {\tt apec} thermal spectrum,
which is given by $\{ 10^{-14} / [ 4 \pi (1+z)^2 D_A^2 ] \} \, \int n_e n_H
\, dV$, where $z$ is the redshift, $D_A$ is the angular diameter distance,
$n_e$ is the electron density, $n_H$ is the ionized hydrogen density,
and $V$ is the volume of the cluster.}
\end{deluxetable*}

%%%%%%%%%%%%%%%%%%%%%%%%%%%%%%%%%%%%%%%%%
\section{CXOU J174437.3-323222: a FSRQ or BL Lac blazar}
\label{sec:blazar}

\begin{deluxetable*}{lcccc}
\tabletypesize{\small}
\tablecaption{Spectral modeling of the \chandras (2008) and \xmms (2012) data for source \#1. \label{tab:s0_cxo-xmm}}
\tablewidth{0pt}
\tablecolumns{5}
\tablehead{ \colhead{Observatory} & \colhead{$N_{\rm H}$}   &  \colhead{$\Gamma$} &\colhead{$\chi^2$ (dof)}  &  \colhead{Flux} \\ 
 & \colhead{(10$^{22}$ cm$^{-2}$)}   & &  &  \colhead{($10^{-12}$ \flux)}
 }
\startdata
\chandras 	 & $3.03_{-0.72}^{+0.91}$	& $1.16_{-0.38}^{+0.42}$	& 1.32 (10 dof)		& $2.60_{-0.39}^{+0.41}$	\\
\xmm & $2.51_{-0.23}^{+0.25}$	& $1.31 \pm 0.09$ 		& 0.89 (397 dof)	& $1.65 \pm 0.07 $		\\
\enddata
%\vspace{-0.8cm}
\tablecomments{The flux in the rightmost column is given for the 0.2--10 keV band, not corrected for absorption. }
\end{deluxetable*}

In this section we focus on source \#1, which has been tentatively identified by \citet{curran.2011fk} as a blazar. We start by comparing the spectra measured by \chandras in 2008 and \xmms in 2012. The spectra from the two instruments are independently fitted with an absorbed power-law model. Table \ref{tab:s0_cxo-xmm} shows that the two spectra are similar within error, but we notice that the 2008 flux is 1.6 times higher than the one measured in 2012.  

\begin{figure}[t]
\begin{center}
\includegraphics[width=0.45\textwidth]{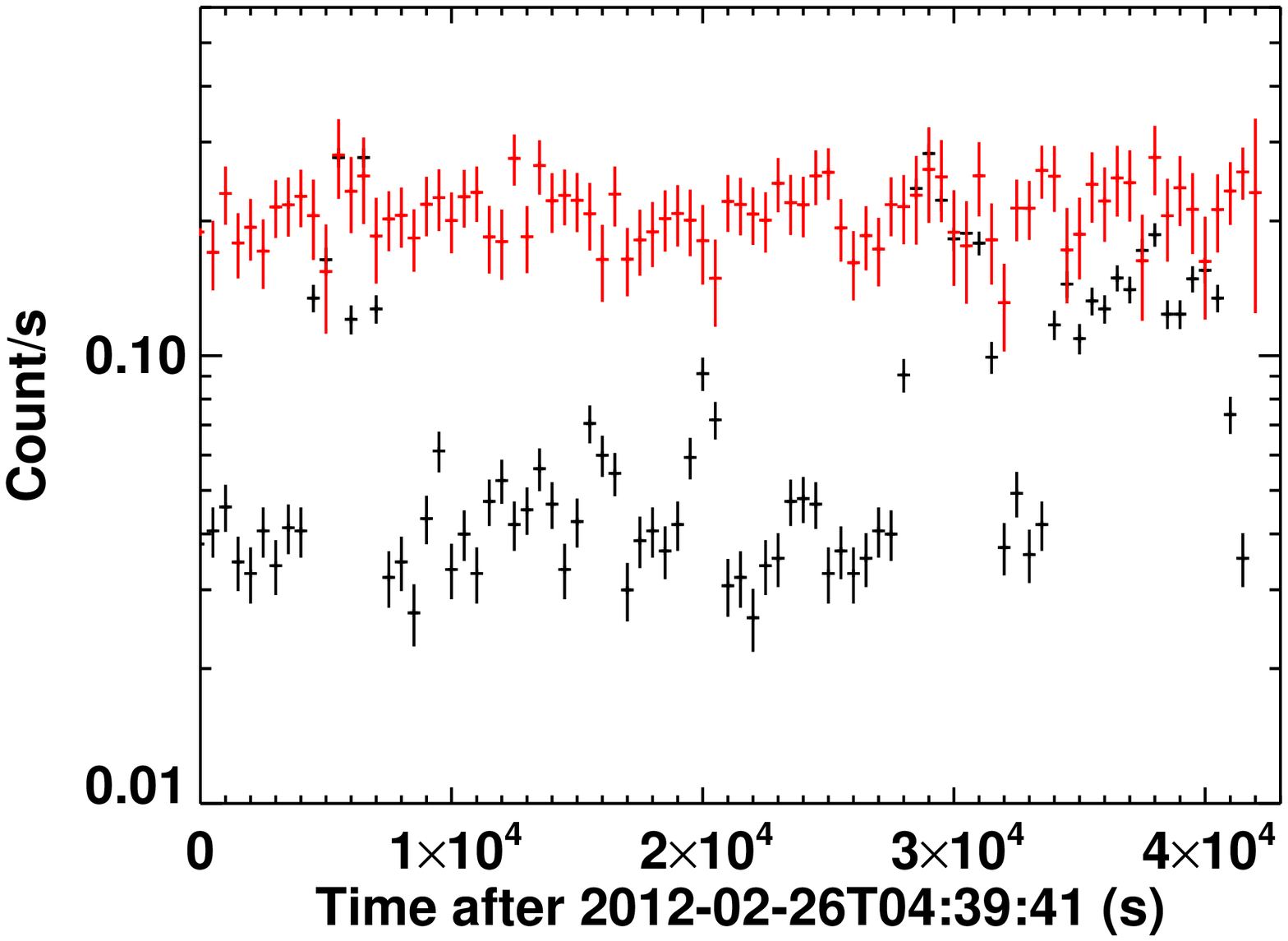}
\caption{\xmms pn 0.4--10 keV light curve (LC) of source \#1 with 500~s binning. The red points show the background-subtracted source LC, and the black points show the background LC (scaled to match the area of the source extraction region), which exhibits several flaring events. The error bars show the 1-$\sigma$ rate confidence limit.
\label{fig:blazarLC}}
\end{center}
\end{figure}

Figure \ref{fig:blazarspc} shows our \xmms spectrum of the blazar candidate along with the \swift-BAT spectrum available publicly from the 70-month catalog\footnote{\url{http://swift.gsfc.nasa.gov/results/bs70mon/}} \citep{baumgartner.2013fk}.  The BAT's angular resolution does not allow the separation between the extended emission and the point source (Figure \ref{fig:image}), so we model the BAT spectrum as the sum of the galaxy cluster ({\tt apec} model) and a power-law. We can see that the power-law becomes dominant above $\approx45$ keV. 
The best joint fit yields a $\chi^2_{\nu}$ of 1.00 for 242 dof. The model includes a cross-normalization constant for each dataset, fixed to 1 for the pn. Their value are $0.80^{+0.05}_{-0.05}$, $0.83^{+0.05}_{-0.05}$ and $0.51^{+0.43}_{-0.37}$ for MOS1, MOS2 and BAT, respectively.  The column density is not affected by the addition of the \swifts data, and is as quoted in Table \ref{tab:s0_cxo-xmm}. The power-law photon index is $1.31 \pm 0.09$, similar to what was found with the \xmms data only (the BAT data alone provide a very poor constraint on the photon index). No cutoff is required by the data, but the points beyond 100~keV should be taken as upper limits. Source \#1's light curve (LC) does not reveal any significant variability (Figure \ref{fig:blazarLC}): the LC can be fit with a constant rate of 0.20 ct/s, returning $\chi^2_{\nu}=0.76$ (83 dof).

\begin{figure}[t]
\begin{center}
\includegraphics[angle=270, width=0.45\textwidth]{f5_s1_xmm+BAT_eeuf_v2.ps}
\caption{Spectrum of source \#1. The black, red and green points are from \xmms MOS1, MOS2, and pn detectors, respectively, while the blue points are from {\it Swift} BAT. The solid lines show the models associated with each dataset, and the bottom panel shows the residuals in units of standard deviation. The blue dotted line shows the decomposition of the model used for the BAT data: the {\tt apec} (from the cluster, source \#0) dominates up to $\approx45$ keV, and the power-law (from the blazar, source \#1) beyond. The error bars show the 1-$\sigma$ confidence level.
\label{fig:blazarspc}}
\end{center}
\end{figure}

\begin{figure}[t]
\begin{center}
\includegraphics[width=0.45\textwidth]{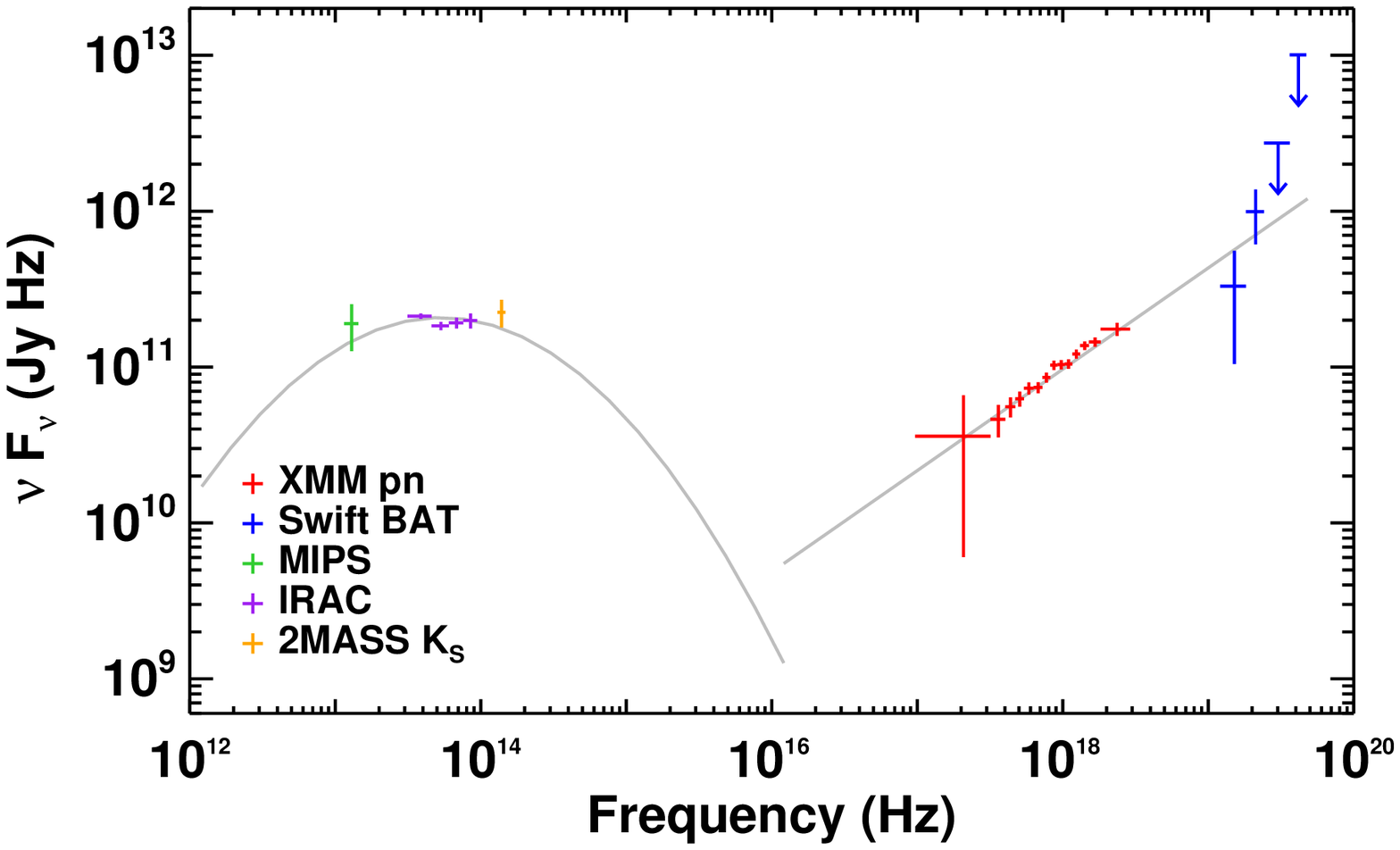}
\caption{Spectral energy distribution of source \#1. The plot shows {\it Spitzer}'s MIPS and IRAC data, 2MASS $K_s$ data point, as well as \xmms pn and \swifts BAT (without the points dominated by the cluster) data in the X-rays. The gray lines illustrate the synchrotron peak and the low-energy tail of the inverse-Compton peak with a non-physical model composed of a log-log parabola in the IR and a power-law in the X-rays.
\label{fig:blazarSED}}
\end{center}
\end{figure}

Figure \ref{fig:blazarSED} shows the spectral energy distribution (SED) of source \#1. Although we are missing a critical part of the spectrum in the optical and UV, this SED matches that of either a flat spectrum radio quasar (FSRQ) or that of a low-frequency peaked BL Lac (LBL) \citep{padovani.1995uq}, which are both described by a synchrotron peak in the $10^{13}$ -- $10^{14}$ Hz range and an inverse Compton peak around $10^{22}$ -- $10^{23}$ Hz. This differs from the analysis proposed by \citet{curran.2011fk} who modeled the IR and X-ray data as a single power-law. Differentiating between an LBL and  a FSRQ requires optical spectroscopy: FSRQs exhibit characteristic broad emission lines in the optical domain, while LBL do not. If this blazar was a FSRQ, the redshift of the emission lines could tell whether this object belongs to the galaxy cluster or if it lies in the background.

We note that change of flux between the \chandras and the \xmms observations is compatible with the blazar scenario. The blazar could have been in outburst during the 2008 observation. %This hypothesis is strengthened by the fact that the {\it Swift}-BAT portion of the spectrum, which is averaged over 70 months, matches well with the projection of the {\it XMM-Newton} SED.

%{\bf looking at the X-shooter data would allow distinguishing between these two types of blazars. However if emissions lines are present, we could find a redshift and tell whether the blazar belongs to the galaxy cluster or lies far behind}.

%observations available: 
%July 2012  run in Paranal we got an X-shooter spectrum (from UV to NIR)  Sylvain Chaty. however it was very faint (taken at high airmass at the end of the observing night
%ISAAC spectrum (Alexis and Sylvain got in July)

%XSHOOTER is a multi wavelength (300-2500nm = 1.2e14 - 1e15 Hz, UV, VIS, NIR) medium resolution spectrograph mounted at the UT2 Cassegrain focus.
%ISAAC is an IR (1 - 5 ?m = 6e13 - 3e14 Hz, IR ) imager and spectrograph that lies at the Nasmyth A focus of UT3

%%%%%%%%%%%%%%%%%%%%%%%%%%%%%%%%%%%%%%%%%
\section{Other sources in the field}
\label{sec:othersrc}
\begin{figure*}[t]
\noindent\includegraphics[width=0.31\textwidth, angle=270]{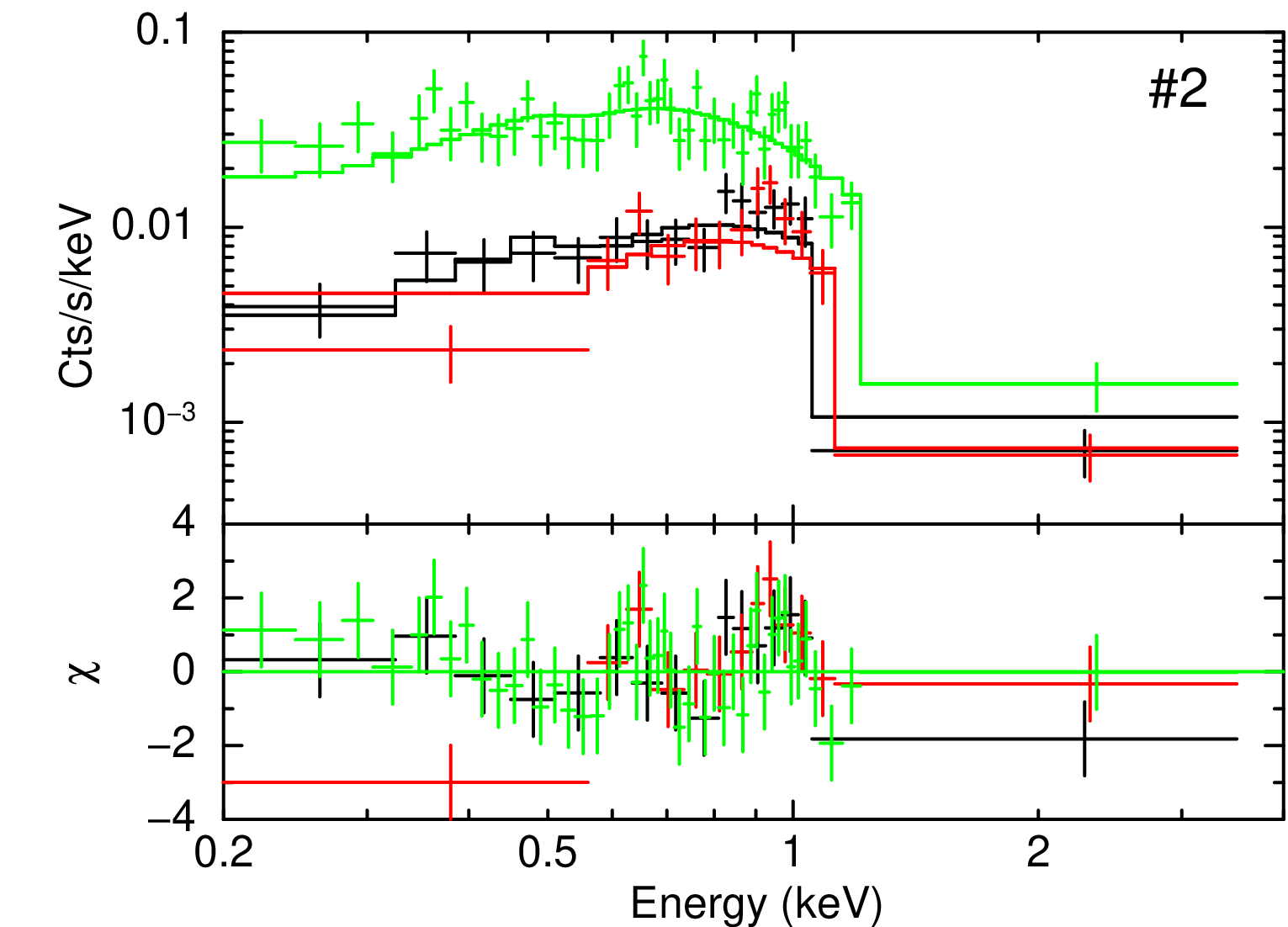}\hfill  %src #2
\noindent\includegraphics[width=0.3\textwidth, angle=270]{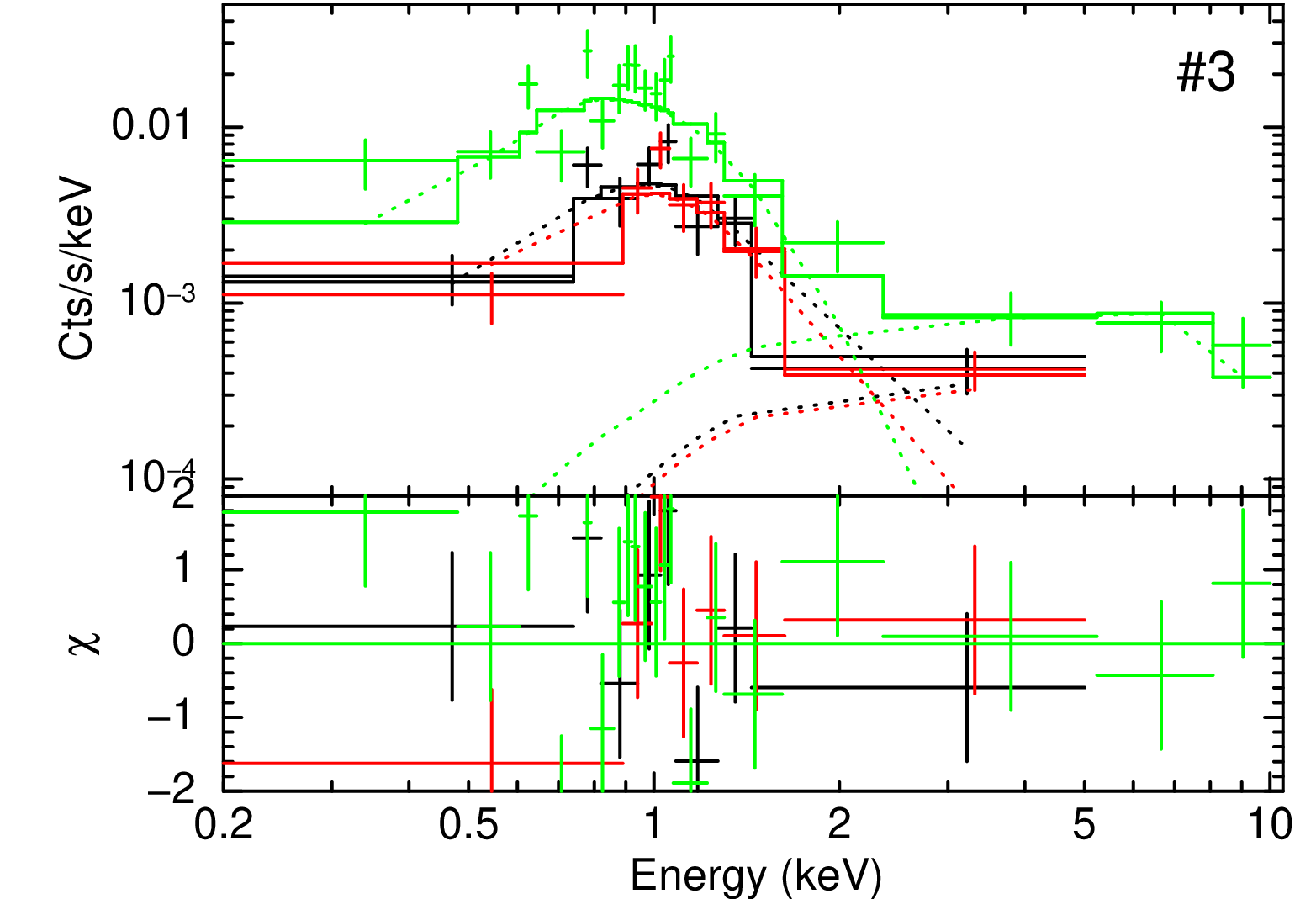}\hfill %src #3
\noindent\includegraphics[width=0.3\textwidth, angle=270]{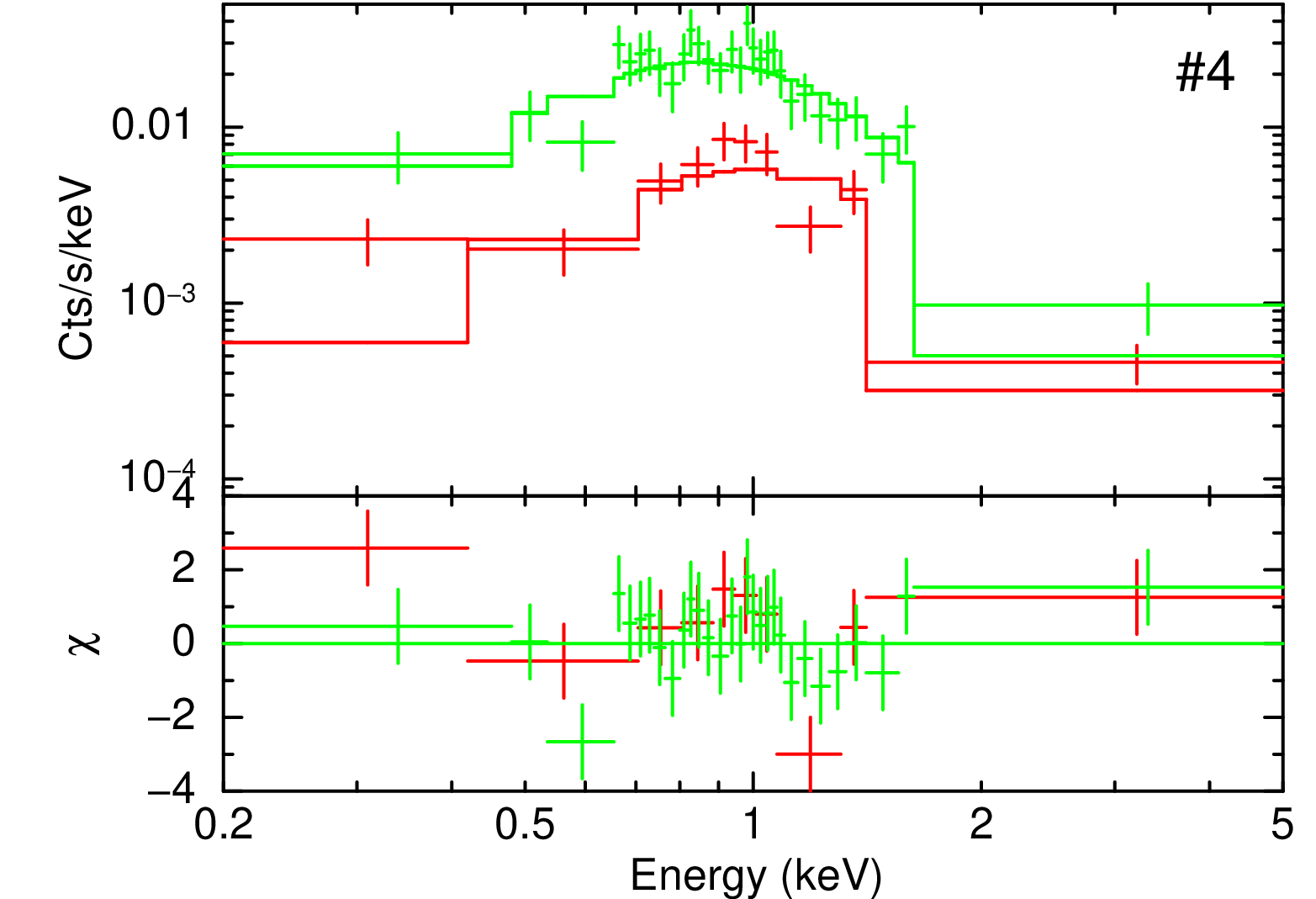}\hfill %src #4
\noindent\includegraphics[width=0.31\textwidth, angle=270]{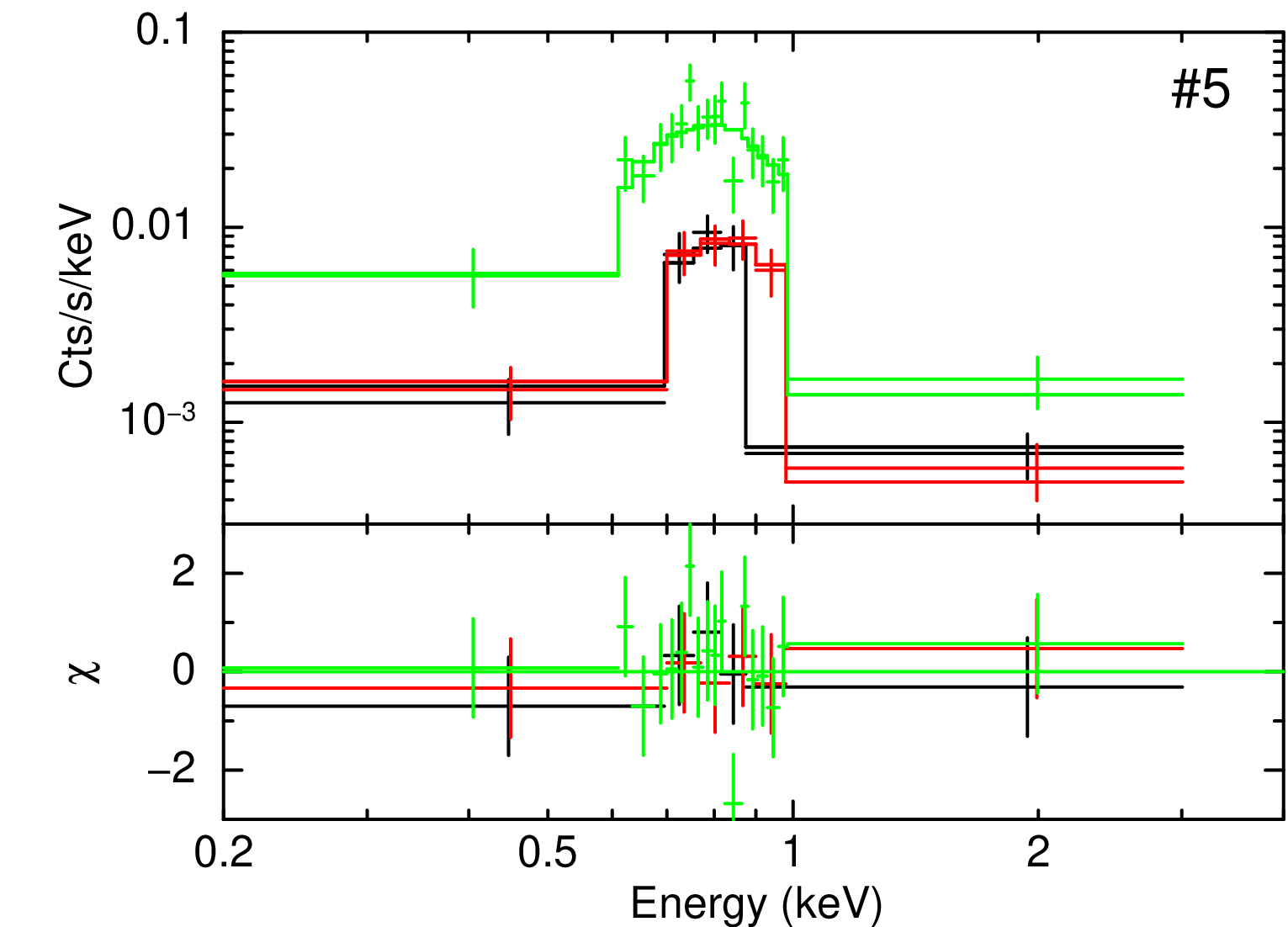} %src #5
\caption{MOS1 (black), MOS2 (red) and pn (green) spectra of sources \#2--5, with the lower panels showing the residuals between model and data in units of standard deviation. Error bars show the $1 \sigma$ confidence range.
\label{fig:ptsrcspc}}
\end{figure*}

Here we focus on sources \#2--5 that are circled in Figure \ref{fig:image}. The goal is to determine whether any of these sources contribute significantly to the flux seen by \inte for \IGR, and whether any might be AGN associated with the cluster. The spectra are extracted from $30''$ radius regions, and they are grouped to yield 40 source + background counts and 3$\sigma$ per bin (including the last bin), the upper energy limit of the last bin being set to maximize its significance. The three spectra are jointly fit with a cross-normalization constant fixed to 1 for pn, and free for MOS1 and MOS2. 
Figure \ref{fig:ptsrcspc} shows the spectra of these sources, and Table \ref{tab:ptsrcfit} presents the number of net counts that was detected in each camera and summarizes their spectral properties. For comparison, the integrated column density in the direction of \IGRs is $\approx7 \times 10^{21}$~cm$^{-2}$ \citep{kalberla.2005uq}. We found no significant variability in the 0.4--10 keV range for any of these sources.

Searching in VizieR\footnote{\url{http://vizier.u-strasbg.fr/viz-bin/VizieR}} at the \chandras positions reported in Table \ref{tab:pointsrc}, we found optical and near infrared counterparts for each of the four sources \citep[as compiled, for instance, in the Naval Observatory Merged Astrometric Dataset, NOMAD;][]{zacharias.2005oq}. We dereddened their $B$, $V$, $J$, $H$ and $K_s$ magnitudes using our X-ray measurement of the column density \citep{guver.2009kx}: $N_{\rm H} ({\rm cm}^{-2}) = (2.21\pm 0.09) \times 10^{21} A_{\rm v}$ (mag), and using the relationship between extinction in the visual band $A_{\rm v}$ and the other bands derived in \citep{cardelli.1989uq}. We then use tables  \citep[e.g. from ][]{2001PhDT..........G,ducati.2001kx,pecaut.2013nx} to match the $B$-$V$, $V$-$J$, $V$-$H$, $V$-$K_s$ color indices (Table \ref{tab:srcmag}) with star spectral types.

\begin{deluxetable}{lcccc}
\tabletypesize{\scriptsize}
\tablecaption{Dereddened magnitudes and color indices for sources 2--5. \label{tab:srcmag}}
\tablewidth{0pt}
\tablecolumns{5}
\tablehead{ \colhead{Source} & \colhead{\#2}   &  \colhead{\#3} &\colhead{\#4}  &  \colhead{\#5} }

\startdata
%$N_{\rm H}$ (10$^{20}$~cm$^{-2}$) 	& $5.20$		& $33.0$		& $20.0$		& $111.0 $	\\
$B$ 	(mag)					& 16.27	 	& 12.57		& 14.080		& 11.542		\\
$V$ 	(mag)					& 15.17		& 12.49		& 13.470		& 11.189		\\
%$R$ 	(mag)				& 14.49		& 12.69		& 13.440		& 10.960		\\
$J$ 	(mag)					& 11.87		& 12.19		& 11.919		& 9.857		\\
$H$	(mag)					& 11.17		& 11.89		& 11.496		& 9.571		\\
$K_s$ (mag)					& 11.03		& 11.85		& 11.369		& 9.450		\\
$B-V$						& 1.09		& 0.09		& 0.31		& -1.34		\\
%$V-R$						& 0.68		& -0.20		& -0.11		& -0.57		\\
$V-J$						& 3.31		& 0.29		& 0.90		& -2.27		\\
$V-H$						& 4.01		& 0.59 		& 1.24 		& -2.45		\\
$V-K_s$						& 4.15		& 0.63		& 1.30		& -2.71		\\
%$J-H$						& 0.70		& 0.30		& 0.34		& -0.18		\\
%$H-K_s$					& 0.14		& 0.04		& 0.06		& -0.26		\\
%Spectral type					& Giant M3	& Supergiant F0 & Supergiant F8 & Unidentified \\
\enddata
%\vspace{-0.8cm}
\end{deluxetable}

%near infrared \citep[2MASS,][]{skrutskie.2006fk}
%mid-infrared \citep[GLIMPSE,][]{benjamin.2003uq,churchwell.2009kx}

%%%%% source 2
\noindent {\bf CXOU J174458.2-323940 (\#2).} We limit the spectral analysis of this source to the 0.2--3.5~keV range, as it has almost no counts beyond 3.5 keV. A thermal model (absorbed black-body) yields a good fit  with $kT=0.19 \pm 0.02$ keV ($\chi^2_{\nu}=1.28$, 71 dof). The very low column density ($N_{\rm H}=5.2^{+4.9}_{-3.6} \times 10^{20}$ cm$^{-2}$) places this source in the foreground. We find a good positional match ($< 0.6''$) with 2MASS~J17445825-3239407, for which the spectral type corresponds to a giant M3 star \citep{ducati.2001kx}. This source could be an active binary star (a binary with two main sequence stars).

%%%%% source 3
\noindent {\bf CXOU J174534.6-322917 (\#3).}  This source's spectral analysis is done over the 0.4--5~keV and 0.4--10~keV for the MOS and the pn data, respectively.  Its spectrum is acceptably fit by the sum of a soft thermal component (black-body with $kT=0.20_{-0.05}^{+0.06}$~keV) and a hard  power-law tail with $\Gamma= -0.68_{-1.02}^{+0.99}$, yielding $\chi^2_{\nu}=1.69$ for 28 dof. The spectrum shows positive residuals around 1 keV, which could be well fit by a Gaussian emission line. Alternatively, replacing the black-body by a  thermal plasma component ({\tt apec}) also improves the fit, although the physical interpretation is not straightforward in either case.

We find a good match with 2MASS~J17453452-3229177: Although it is $1.76''$ away, the inspection of the 2MASS images show that the \chandras error circle ($4.15''$ radius for this source) nicely overlaps with the IR source without any possible confusion. The colors of this counterpart are not well matched by any type of main sequence star, however they coincide well with the supergiant type F0 \citep{ducati.2001kx}. Assuming that the black-body + power-law model is correct for the \xmms spectrum, source \#3 could be an X-ray binary.

%%%%% source 4
\noindent {\bf CXOU J174510.9-322655 (\#4).} The spectral analysis of source \#4 is done over the 0.2--5 keV energy range. Its spectrum is best fit by an absorbed power-law with $\Gamma=4.66_{-0.85}^{+1.14}$ ($\chi^2_{\nu}=1.23$, 36 dof), however such a soft spectrum is physically better interpreted as thermal emission. Using an absorbed black-body model, we find $N_{\rm H}=2.0_{-1.2}^{+1.6} \times 10^{21}$ cm$^{-2}$ and $kT = 0.24_{-0.04}^{+0.05}$~keV  ($\chi^2_{\nu}=1.42$, 36 dof). We find a counterpart located $0.42''$ away from the \chandras position, 2MASS~J17451093-3226560, which color indices match well with a F8 supergiant \citep{ducati.2001kx}, or with a F--G main sequence star. Similarly to source \#2, this source could be an active binary star.

%%%%% source 5
\noindent {\bf CXOU J174428.4-322828 (\#5).} The spectrum of source \#5 is very soft; we perform the spectral analysis over the 0.2--3 keV range. The model yielding the best fit is an absorbed black-body, with $kT=8.2^{+1.6}_{-1.8} \times 10^{-2}$ keV ($\chi^2_{\nu}=0.81$, 24 dof).  2MASS~J17442842-3228286 is a robust counterpart, $0.06''$ away from the \chandras position. The NIR and optical photometry indicate a spectrum rising towards shorter wavelength, possibly connecting with the tail of the black body spectrum observed in X-rays. This object might be an isolated neutron star, although one would need to build a SED of the source to confirm this hypothesis, which goes beyond the scope of this paper. In any case, it is not an AGN.

%%% Conclusion of this section
With its hard power-law tail, source \#3 is the only one that is hard enough to contribute to \IGR. However, with an order of magnitude lower flux  than the blazar (source \#1), and no indication of variability, its contribution is not likely to be significant.

\hfill\null
\noindent\null\hfill

%%%%%%%%%%%%%%%%%%
\section{Summary}
This observation revealed that the extended object in the field of \IGRs is a massive, nearby galaxy cluster heretofore hidden by the plane of the Galaxy, the Scorpius cluster. Since extragalactic surveys avoid the plane and nearly all the cluster emission within the ROSAT passbands has been absorbed by intervening cold gas, such a serendipitous discovery is not unexpected. The larger effective area, especially at harder energies, makes \xmms the ideal X-ray observatory for revealing the nature of this cluster without supporting information at other wavelengths.

Although treated as a symmetric cluster to determine its gross properties, the surface brightness contours in Figure~\ref{fig:clusterimages} illustrate asymmetric features that suggest a disturbed intracluster medium. If these features are due to a recent major merger, then we expect temperature variations to correlate with the surface brightness structures. Preliminary work suggests that this is indeed the case, but a more in-depth spatial-spectral analysis is beyond the scope of the current paper. Because slight spatial variations in Galactic foreground and absorption can bias local temperature estimates, the spatially flat values assumed here are insufficient to accurately determine the features of this new cluster. A more detailed analysis will be presented in a later paper (Wik et al. in prep).

As illustrated in Figure \ref{fig:blazarspc} by the \swift-BAT data, \IGRs is actually 2 sources: it is dominated in the 20--45~keV band by the galaxy cluster and in the 45-100 keV by the blazar. We confirmed the tentative identification by \citet{curran.2011fk} of CXOU J174437.3-323222 being a blazar (although we interpret its SED differently), and we propose that it is either a FSRQ or a low-frequency peaked BL Lac. We also analyzed four other sources present in the field of this \xmms observation, but we concluded that they are most likely Galactic sources (foreground) that did not contribute to \IGR.

\acknowledgments
Partial support for this work was provided by NASA through Astrophysics Data Analysis Program (ADAP)
grant NNX12AE71G. The authors are grateful to Marco Ajello and Roman Krivonos for useful discussions.

\begin{deluxetable*}{cccccccc}
 \tabletypesize{\scriptsize}% \tabletypesize{\footnotesize }
\tablecaption{Net counts, best fit model parameters, and flux for sources \#2--5 \label{tab:ptsrcfit}}
\tablecolumns{8}
\tablewidth{0.85\textwidth}
\tablehead{ \colhead{ID \#} & \colhead{MOS1}   &  \colhead{MOS2 }  &  \colhead{pn} &  \colhead{$kT$} & \colhead{$\Gamma$} &\colhead{$N_{\rm H}$ }  & \colhead{Flux}  \\
  & (cts) & (cts) & (cts) & (keV) & &($10^{21}$ cm$^{-2}$) &  (erg/s/cm$^2$) }
\startdata
2 & $394.2 \pm 29.1$	& $334.1 \pm  28.2$	& $1231.2 \pm 56.1$	& $0.19 \pm 0.02$ 		& \centering{--}			& $0.52_{-0.36}^{+0.49} $		& $7.1_{-0.58}^{+0.63} \times 10^{-14}$  \\
3 & $239.4 \pm 23.7$	& $188.9 \pm 21.3$	& $803.9 \pm  59.2$	& $0.20_{-0.05}^{+0.06}$	& $ -0.68_{-1.02}^{+0.99}$ & $3.3_{-2.1}^{+3.3} $		& $2.3_{-0.8}^{+0.9} \times 10^{-13}$ \\
4 & \centering{--}		& $272.8 \pm 25.0$	& $864.3 \pm 57.6$	& $0.24_{-0.04}^{+0.05}$	&  \centering{--}			& $2.0_{-1.2}^{+1.6} $		& $5.3_{-0.68}^{+0.78} \times 10^{-14}$\\
5 & $154.0 \pm  20.5$	&$173.7 \pm  21.6$	& $578.5 \pm  51.7$	& $0.081\pm0.017$		&  \centering{--}			& $11.1^{+4.6}_{-2.8}$		& $3.0_{-0.39}^{+0.41} \times 10^{-14} $\\
\enddata
\vspace{-0.4cm}
\tablecomments{The flux in the rightmost column is given for the 0.2--10 keV band, not corrected for absorption. }

\end{deluxetable*}

\clearpage


%% in parentheses, after the keyword, but they will not be verified.

{\it Facilities:}  \facility{XMM}, \facility{INTEGRAL (IBIS)}, \facility{CXO (ACIS)}.



%\bibliographystyle{jwapjbib}
\bibliography{IGRJ17448}

\clearpage

\end{document}